\documentclass{PoS}
\usepackage{graphicx}
\ifpdf
\usepackage{epstopdf}
\fi
\usepackage[multidot]{grffile}
\usepackage{fixltx2e}

\usepackage[latin1]{inputenc}
\usepackage{graphics}
\usepackage{graphicx}
\usepackage{epsfig}
\usepackage{color}
\usepackage{longtable}
\usepackage{latexsym}
\usepackage{amsmath}
\usepackage{amsthm}
\usepackage{amsfonts}
\usepackage{amssymb}
\usepackage{dsfont}
\usepackage{bm}
\usepackage{graphics,psfrag}
\usepackage{graphicx,psfrag}
\usepackage[nice]{nicefrac}

\newcommand{\be}{\begin{equation}}
\newcommand{\ee}{\end{equation}}
\newcommand{\bea}{\begin{eqnarray}}
\newcommand{\eea}{\end{eqnarray}}

\renewcommand{\imath}{\mathrm{i}}

\newlength{\mylenC}
\newlength{\mylenL}
\newlength{\mylenR}
\newlength{\mylenRm}
\title{Excited light and strange hadrons from the lattice with two Chirally Improved quarks}

\ShortTitle{Excited light and strange hadrons from the lattice with two Chirally Improved quarks}

\author{\speaker{Georg P.~Engel} 
\\
Universit\`a Milano-Bicocca, Italy and \\
INFN, Sezione di Milano-Bicocca, Italy \\
E-mail: \email{georg.engel@mib.infn.it}}

\author{C.B.~Lang\\
Institut f\"ur Physik, FB Theoretische Physik, \\ Universit\"at Graz, A--8010 Graz, Austria\\
E-mail: \email{christian.lang@uni-graz.at}}
\author{Daniel Mohler\\
Fermi National Accelerator Laboratory, Batavia, Illinois 60510-5011, USA \\        
E-mail: \email{dmohler@fnal.gov}}
\author{Andreas Sch\"afer\\
Institut f\"ur Theoretische Physik, \\ Universit\"at Regensburg, D--93040 Regensburg, Germany\\
E-mail: \email{andreas.schaefer@physik.uni-regensburg.de}}

\abstract{Results for excited light and strange hadrons from the lattice with two flavors of Chirally Improved sea quarks are presented. We perform simulations at several values of the pion mass ranging from 250 to 600 MeV and extrapolate to the physical pion mass.  
The variational method is applied to extract excited energy levels but also to discuss the content of the states. Among others, we explore the flavor singlet/octet content of Lambda states. In general, our results agree well with experiment, in particular we confirm the Lambda(1405) and its dominant flavor singlet structure. 
}

\FullConference{XV International Conference on Hadron Spectroscopy-Hadron 2013\\
		4-8 November 2013\\
		Nara, Japan }

\begin{document}
%%%%%%%%%%%%%%%%%%%%%%%%%%%%%%%%%%%%%%%%%%%%%%%%%%
%%%%%%%%%%%%%%%%%%%%%%%%%%%%%%%%%%%%%%%%%%%%%%%%%%

%%%%%%%%%%%%%%%%%%%%%%%%%%%%%%%%%%%%%%%%%%%%%%%%%%
\section{Introduction}
%%%%%%%%%%%%%%%%%%%%%%%%%%%%%%%%%%%%%%%%%%%%%%%%%%
\noindent
Our main knowledge about Strong Interactions lies in experimental data on hadron resonances \cite{Beringer:1900zz},  
of which an {\it ab-initio} determination starting from QCD would be truly desirable. 
Although there is noticeable progress in computing resonance properties on the lattice (for a review, see, e.g., \cite{Mohler:2012nh}), 
this is still a prohibitively difficult task for most resonances. 
To discuss an extensive list of excited hadrons, we consider the discrete spectrum of the Hamiltonian in a finite box.
This discrete spectrum becomes denser towards larger volumes, and the volume dependence is related to the phase shift of the resonance in the elastic region \cite{Luscher:1990ux,Luscher:1991cf}.
However, in finite volume and also for unphysically heavy pion masses the decay channels are often closed or the related phase space is small. 
Correspondingly, the energy levels in the finite system are close to the resonance peak, in particular for narrow resonances. 
Hence, as a first approximation, the discrete energy levels can be identified with the masses of corresponding resonances. 
We discuss all canonical channels of isovector light and strange mesons and light and strange baryons and give results at the physical pion mass which can be compared to experiment. 
A large basis of interpolators is considered, which, however, includes only quark--antiquark and 3-quark interpolators. 
In principle, the sea quarks should provide overlap of these interpolators with meson--meson and meson--baryon states.
In practice, however, we cannot clearly identify such states.
A possible explanation may be a weak overlap with the used interpolators, indicating the need for more general interpolators for future work. 
The results presented here have been published before in \cite{Engel:2011aa,Engel:2011pp,Engel:2012qp,Engel:2013ig,Engel:2013ita}, 
for recent reviews on hadron spectroscopy on the lattice, see, e.g., \cite{Bulava:2011np,Lin:2011ti,Fodor:2012gf}. 

%%%%%%%%%%%%%%%%%%%%%%%%%%%%%%%%%%%%%%%%%%%%%%%%%%
\section{Setup}
%%%%%%%%%%%%%%%%%%%%%%%%%%%%%%%%%%%%%%%%%%%%%%%%%%
\noindent
While most lattice Dirac operators break chiral symmetry explicitly, a lattice version of the symmetry can be formulated by choosing a particular discretization of the Dirac operator, which obeys the non-linear ``Ginsparg-Wilson" equation \cite{Ginsparg:1981bj,Luscher:1998pqa}.
We use the Chirally Improved (CI) fermion action \cite{Gattringer:2000js,Gattringer:2000qu}, which is an approximate solution to this equation.
For the gauge sector we use the tadpole-improved L\"uscher-Weisz action \cite{Luscher:1984xn}. 
The lattice spacing $a$ is set at the physical pion mass for each value of the gauge coupling using a Sommer parameter $r_0=0.48$ fm, as discussed in \cite{Engel:2011aa}.
The lowest energy levels can be extracted considering correlation functions of hadronic interpolators and their exponential decay in euclidean time. 
The correlators are computed using Monte Carlo simulations with importance sampling for the gauge sector and the fermion determinant. 
To discuss excited states, the variational method is applied \cite{Luscher:1990ck,Michael:1985ne}. 
One constructs several interpolators $O_i$ for each set of quantum numbers 
and computes the cross-correlation matrix $C_{ij}(t)=\langle 
O_i(t) O_j(0)^\dagger\rangle$. 
Its generalized eigenvalue problem
yields approximations to the eigenstates of the Hamiltonian. 
The exponential decay of the eigenvalues 
is governed by the energy levels, and the eigenvectors tell about the content of the states 
in terms of the lattice interpolators. 
We simulate two CI light sea quarks and consider a valence strange quark. 
We generate seven ensembles 
with pion masses in the range from 250 to 600 MeV, 
lattice spacings between 0.13 and 0.14 fm and lattices of size $L\approx2.2$ fm. For two ensembles with light pion masses
also lattices with different volumes are considered to discuss finite volume effects. 
The interpolators are given explicitly in \cite{Engel:2011aa,Engel:2013ig}.
The present work is a continuation of \cite{Engel:2010my}, considering more ensembles, larger statistics and more observables. 

%%%%%%%%%%%%%%%%%%%%%%%%%%%%%%%%%%%%%%%%%%%%%%%%%%
\section{Results}
%%%%%%%%%%%%%%%%%%%%%%%%%%%%%%%%%%%%%%%%%%%%%%%%%%
%
\begin{figure}[htb]
\centerline{
\noindent\includegraphics[width=\mylenL,clip]{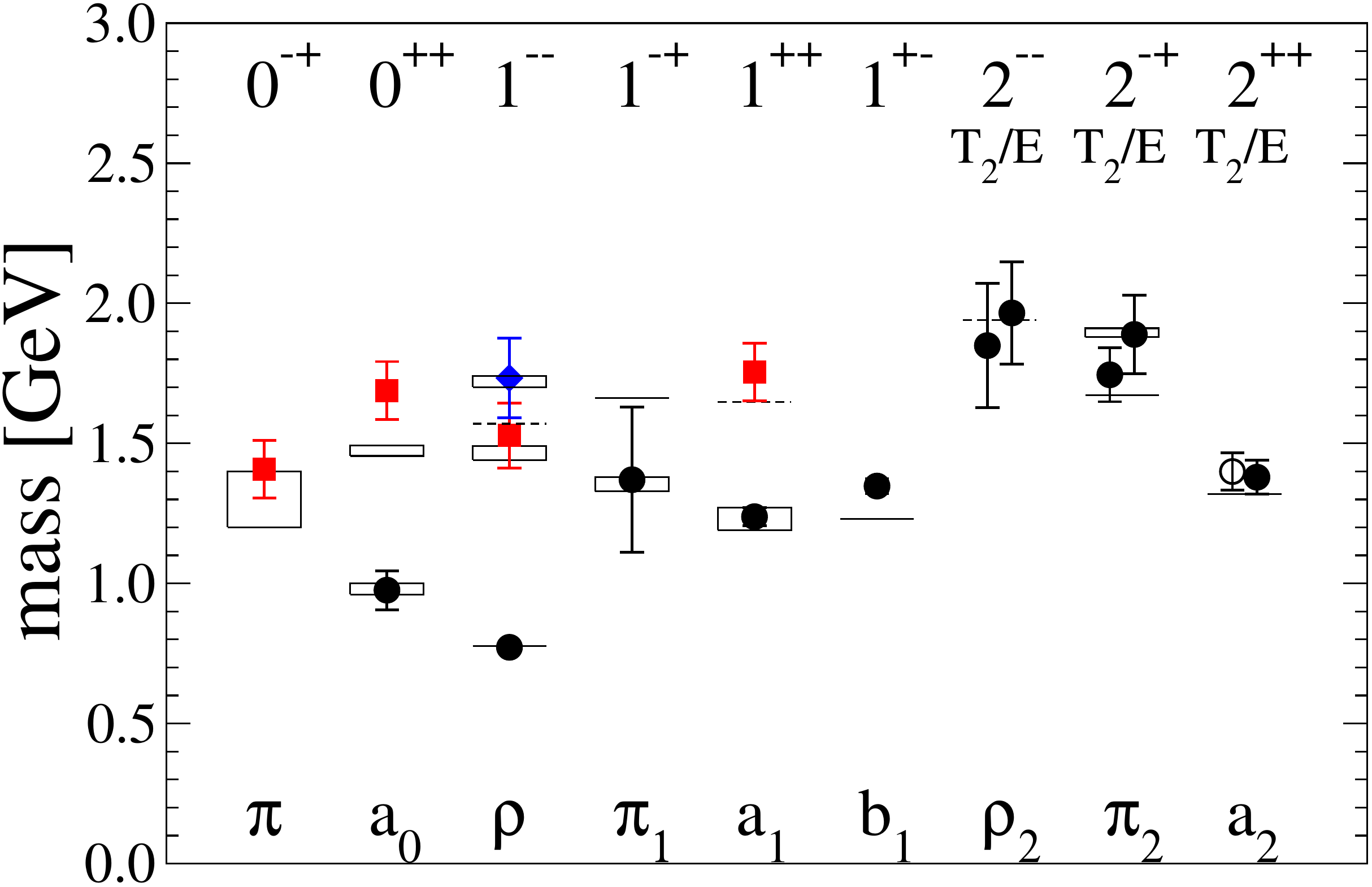}
\noindent\includegraphics[width=\mylenR,clip]{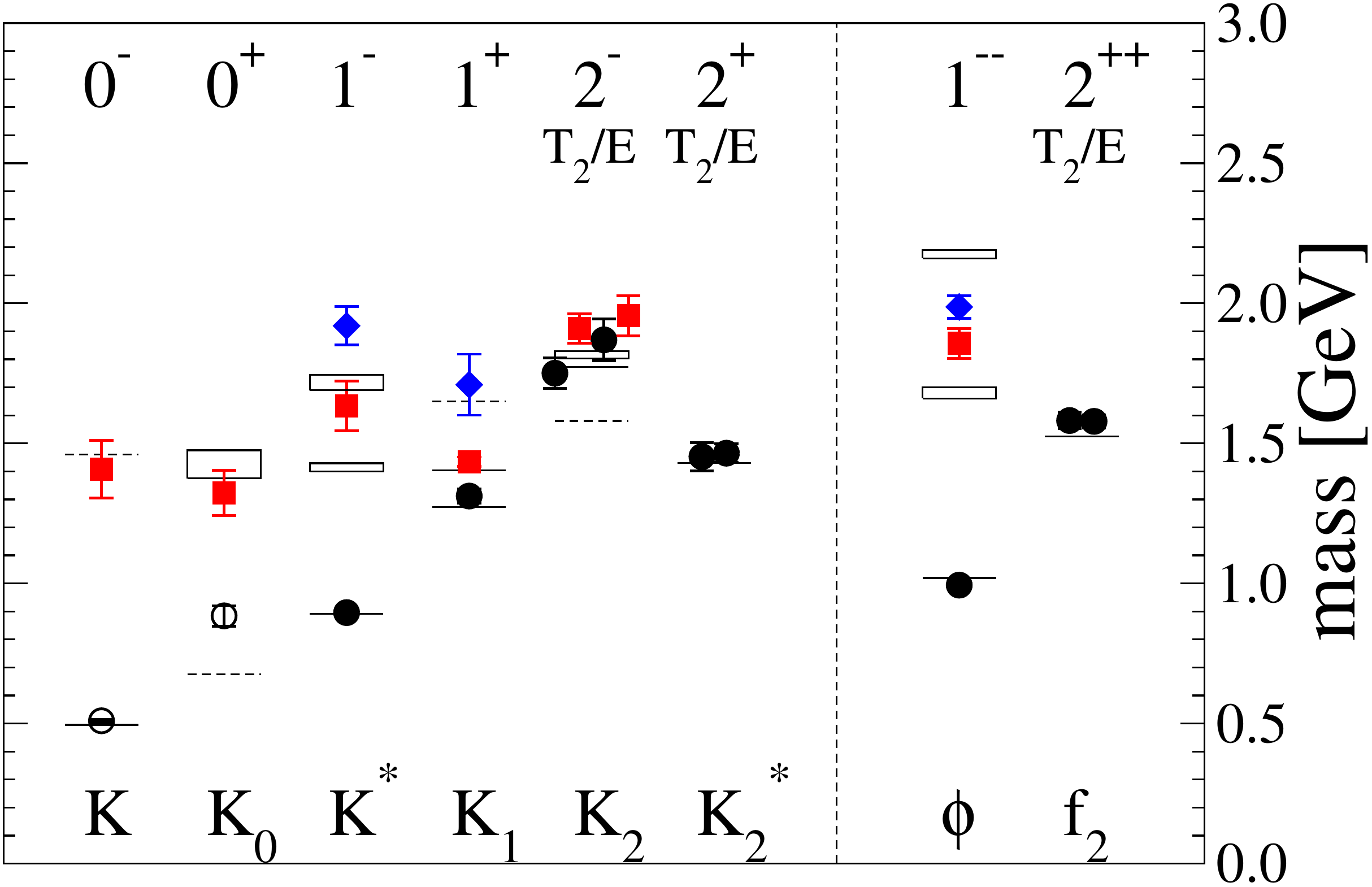}
}
\caption{
Energy levels for isovector light (left), strange and isoscalar (right) mesons in finite volume ($L\approx2.2$ fm).
All values are obtained by extrapolation linear in the pion mass squared to the physical pion mass.
Horizontal lines and boxes represent experimentally known states, 
dashed lines indicate poor evidence, according to \cite{Beringer:1900zz}.
The statistical uncertainty of our results is indicated by bands of $1\sigma$, 
that of the experimental values by boxes of $1\sigma$.
For spin 2 mesons, results for $T_2$ and $E$ are shown side by side.
Open symbols denote a poor $\chi^2$/d.o.f.~of the chiral fits. 
Disconnected diagrams  are neglected in the isoscalar channels.
}
\label{fig:mesons_summary}
\end{figure}
\begin{figure}[htb]
\vspace{-4mm}
\centerline{
\noindent\includegraphics[width=\mylenL,clip]{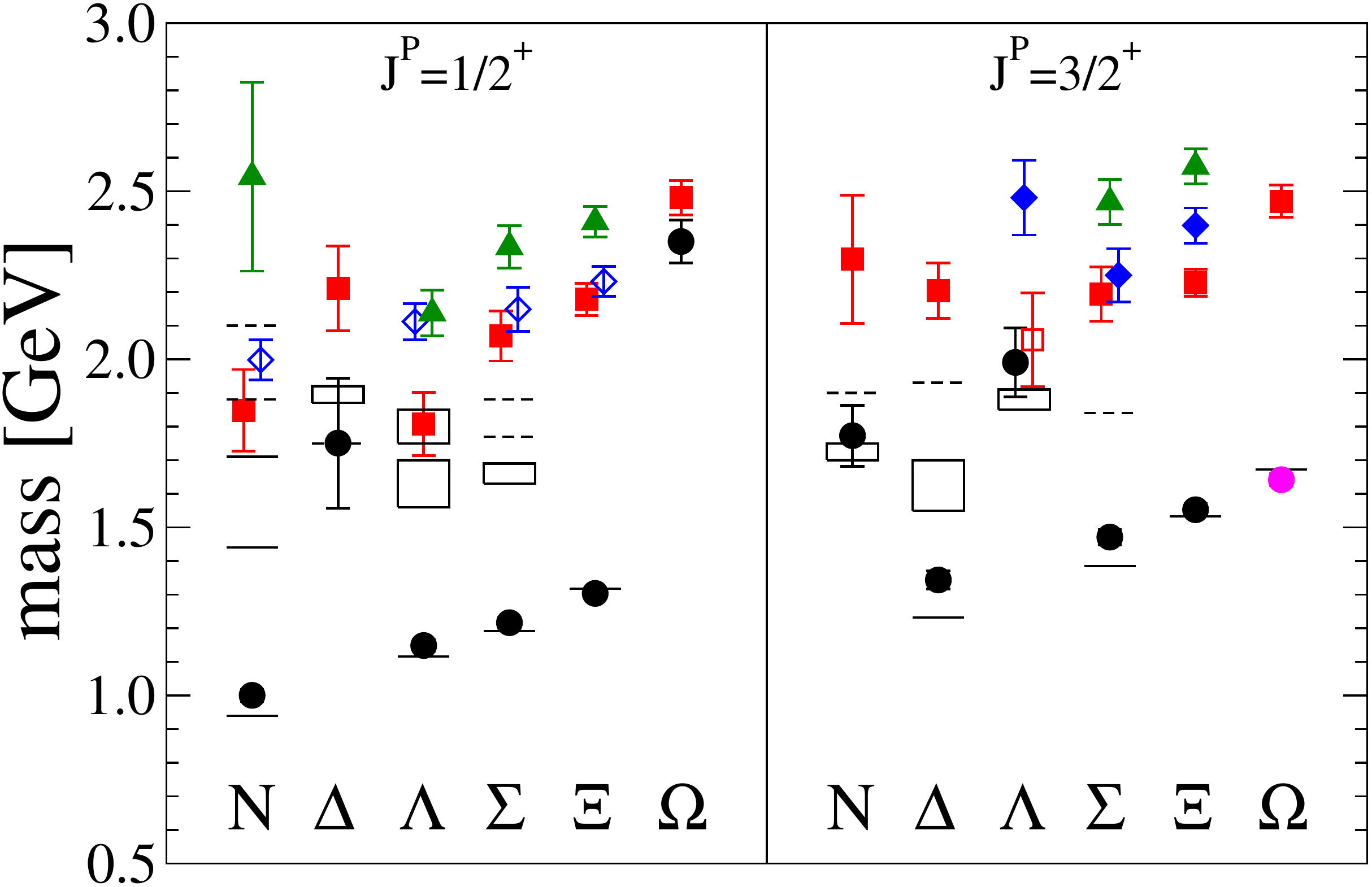}
\noindent\includegraphics[width=\mylenR,clip]{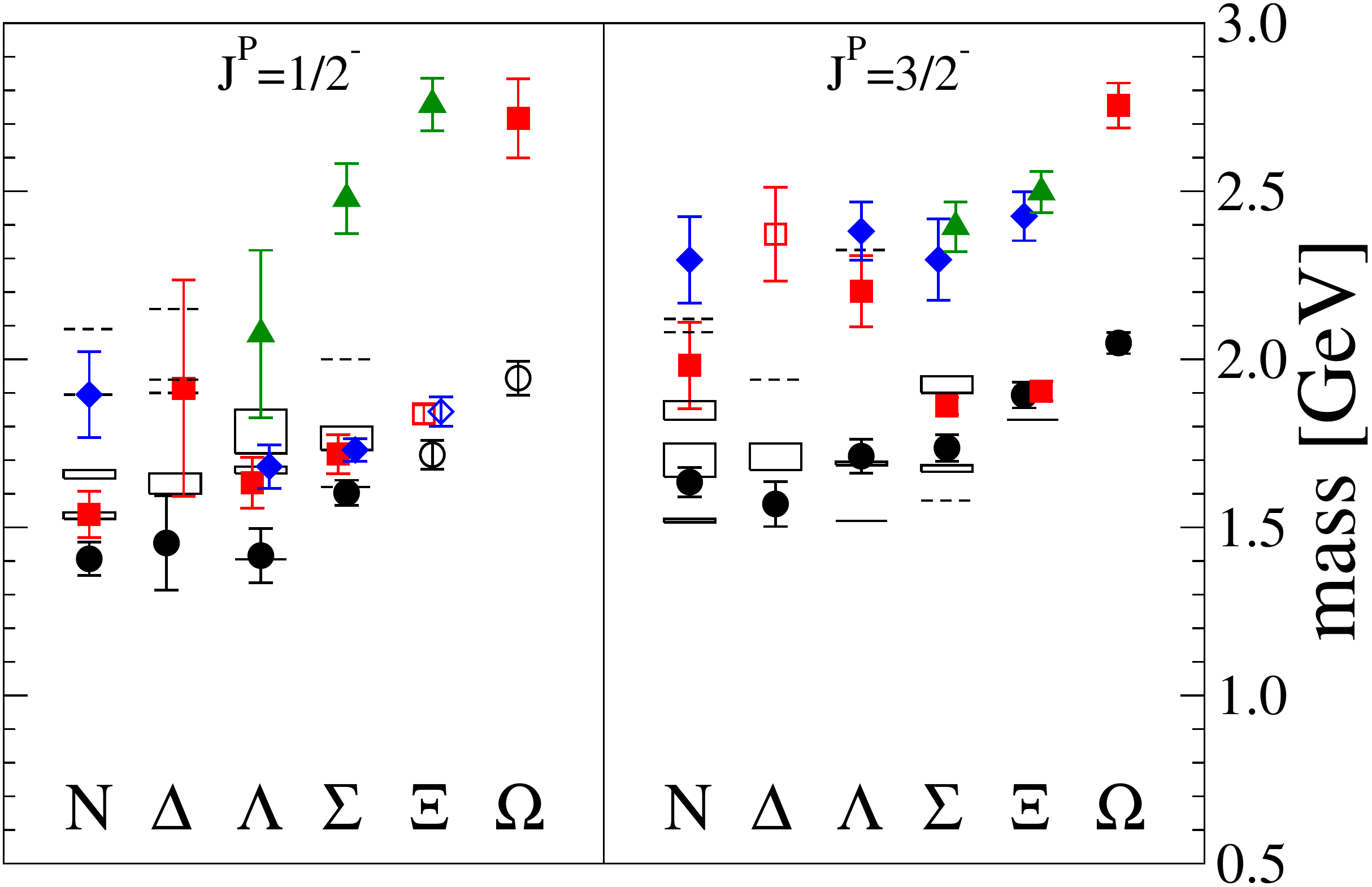}
}
\vspace{-5mm}
\caption{
Like Fig.~1, 
but for positive (left) and negative parity (right) baryons.}
\label{fig:baryons_summary}
\end{figure}
\begin{figure}[htb]
\centerline{
\hspace{-1mm}
\begin{minipage}[c]{0.48\textwidth}
\vspace{-10mm}
\includegraphics[width=\mylenR,clip]{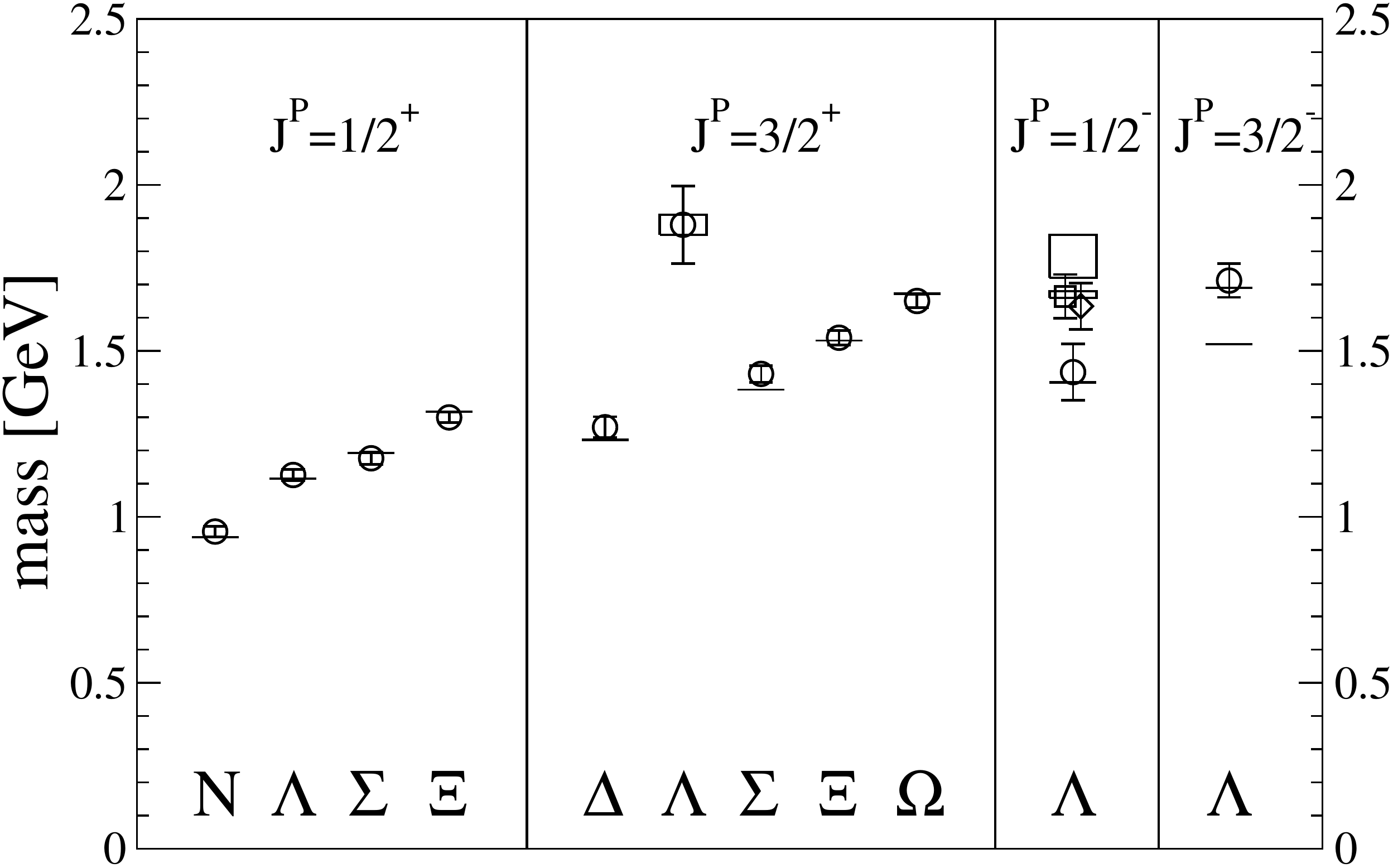}
\end{minipage}
\hspace{1mm}
\begin{minipage}[c]{0.48\textwidth}
\vspace{-3mm}
\includegraphics[width=\mylenL,clip]{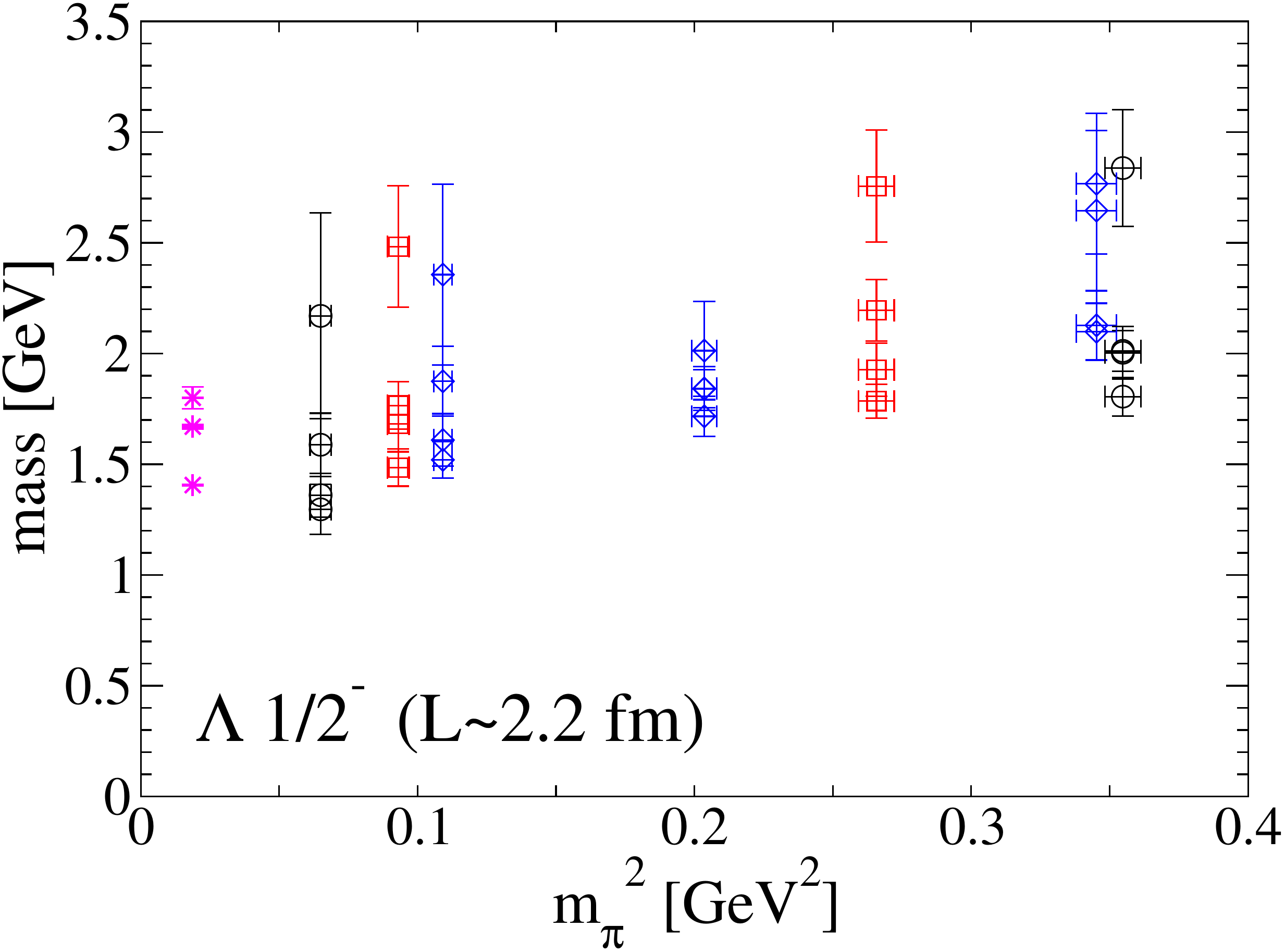}
\end{minipage}
}
\caption{
Lhs: Like Fig.~1, 
but for hadrons in the infinite volume limit.
Rhs: Energy levels for the baryon channel $\Lambda$ ($J^P=1/2^-$) in a finite box of linear size $L\approx 2.2$ fm.
The magenta stars denote the experimental values \cite{Beringer:1900zz}, all other symbols (at heavier pion masses) represent results from the simulations.  
}
\label{fig:infvol_summary}
\end{figure}
\begin{figure}
\vspace{-4mm}
\centerline{
\hspace{0mm}
\noindent\includegraphics[width=0.4\textwidth,angle=-90,clip]{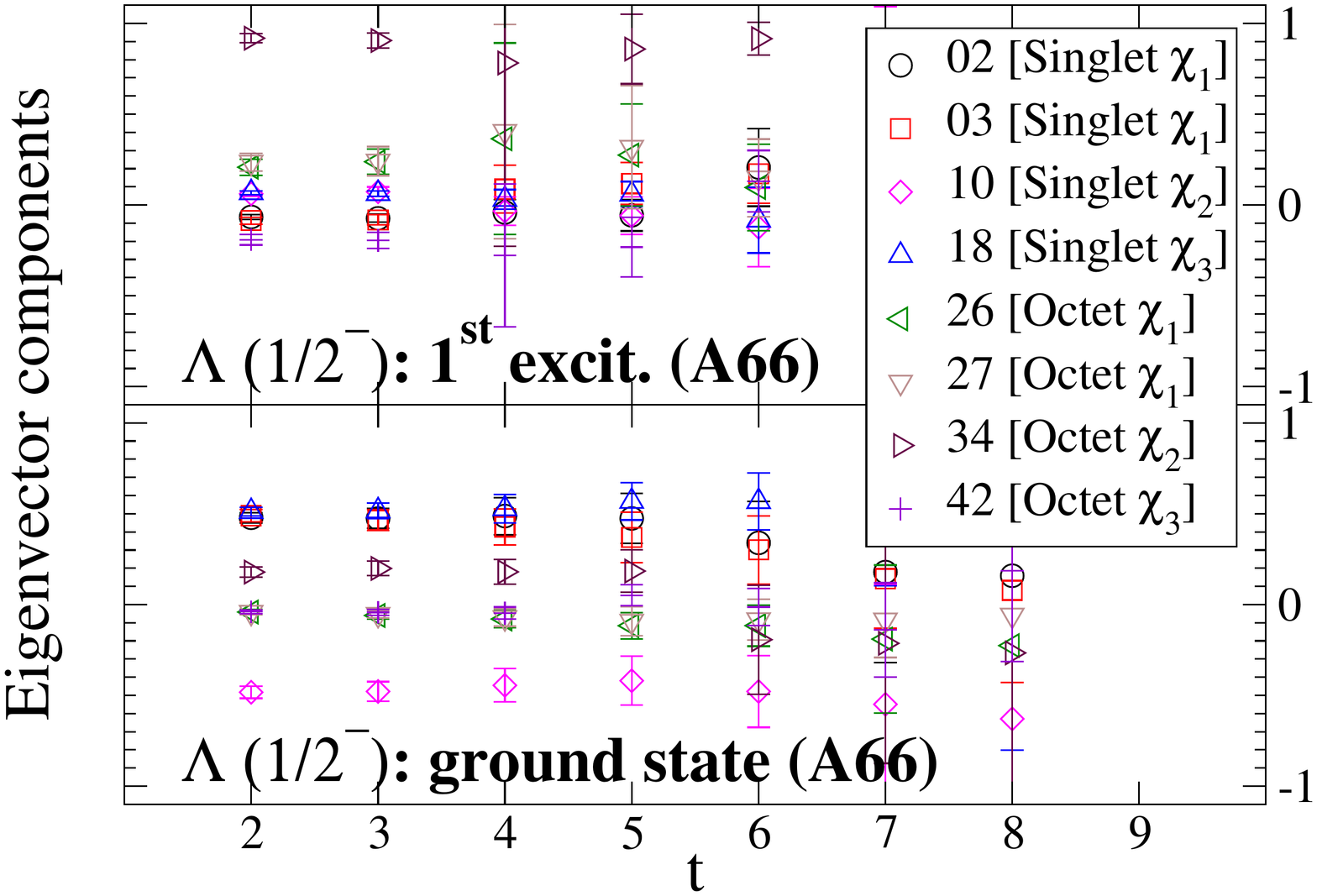}
\hspace{-10mm}
\noindent\includegraphics[width=0.4\textwidth,angle=-90,clip]{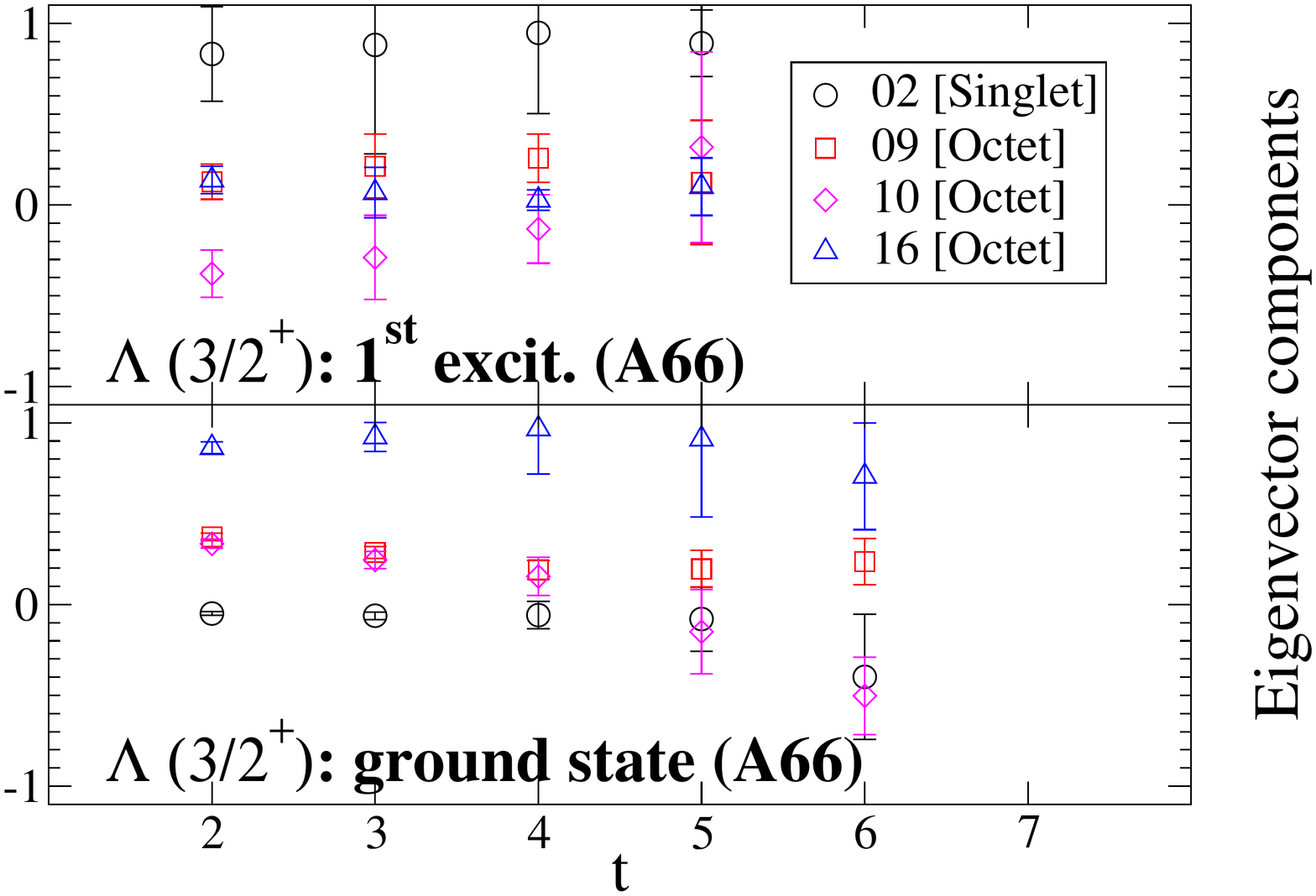}
}
\vspace{-4mm}
\caption{
Lhs: Eigenvectors for the ground state and the first excitation for $m_\pi\approx 255$ MeV,  for the baryon channel $\Lambda$ ($J^P=1/2^-$). The ground state $\Lambda(1405)$ is dominated by flavor singlet, the first excitation by octet interpolators. 
Rhs: Same as left hand side, but for $J^P=3/2^+$. 
We find the first excitation to be dominated by flavor singlet interpolators, which vanish exactly for point-like interpolators due to Fierz-identities. 
}
\label{fig:lambda}
\end{figure}

We study the isovector light and strange meson channels $J=0,1,2$ in both parities and both $C$-parities and the light and strange baryon channels $J=1/2,3/2$ in both parities. 
Figs.~\ref{fig:mesons_summary} and \ref{fig:baryons_summary} show our results for the energy levels in finite volume after extrapolation to the physical pion mass.
In general, the results agree nicely with the experimental values, where available. 
As an example, we reproduce the prominent $\Lambda(1405)$. 
In Fig.~\ref{fig:infvol_summary} (rhs) we show the corresponding
results for the individual ensembles in this channel.
In some cases our results suggest  the existence of yet experimentally unobserved resonance states. 
However, some obtained energy levels mismatch experiment. 
The first excitation in the nucleon ($J^P=1/2^+$) channel, e.g., lies considerably
higher than the Roper resonance. A possible interpretation is
a weak overlap of our interpolators with the physical state. 
Another possible source of deviation from experiment are finite-volume effects. 
Fig.~\ref{fig:infvol_summary} shows our results for a selection of states after extrapolation to infinite volume. 
In general our results in the infinite volume limit compare very well with experiment.
For the baryons, we analyze the flavor content by identifying the singlet/octet/decuplet contributions. 
For example, we find a dominance of singlet interpolators (mixing of 15-20\% with octet) for $\Lambda(1405)$, 
and a dominance of octet interpolators of the first excitation.
The corresponding eigenvectors are shown in Fig.~\ref{fig:lambda}. 
In the $\Lambda$ ($J^P=3/2^+$) channel, the first excitation appears to be dominated by flavor singlet interpolators. 
Such interpolators vanish exactly for point-like quark-fields due to Fierz-identities. 
Using different quark-smearing widths to side-step these identities, we are able to construct non-vanishing interpolators nevertheless. 
To conclude, we remark that our {\it ab-initio} determination of the excited hadron spectrum agrees well with experiment, and provides also further insights, such as yet unknown states, 
and also the content of the states in terms of lattice interpolators. 

%%%%%%%%%%%%%%%%%%%%%%%%%%%%%%%%%%%%%%%%%%%%%%%%%%
\section{Acknowledgments}
%%%%%%%%%%%%%%%%%%%%%%%%%%%%%%%%%%%%%%%%%%%%%%%%%%
\noindent
We thank E.~Gamiz, Ch.~Gattringer, L.~Y.~Glozman, M.~Limmer, 
W. Plessas, H.~Sanchis-Alepuz, M.~Schr\"ock and V.~Verduci
for valuable discussions.  The calculations have been performed at the Leibniz-Rechenzentrum Munich and on clusters at the
University of Graz. We thank these institutions.
G.P.E.~and A.S.~acknowledge support by
the DFG project SFB/TR-55. 
G.P.E.~was supported by the MIUR--PRIN 20093BM-NPR. 
Fermilab is operated by Fermi Research Alliance, LLC under Contract No. De-AC02-07CH11359 with the United States Department of Energy.

%%%%%%%%%%%%%%%%%%%%%%%%%%%%%%%%%%%%%%%%%%%%%%%%%%

%%%%%%%%%%%%%%%%%%%%%%%%%%%%%%%%%%%%%%%%%%%%%%%%%%

%%%%%%%%%%%%%%%%%%%%%%%%%%%%%%%%%%%%%%%%%%%%%%%%%%
%%%%%%%%%%%%%%%%%%%%%%%%%%%%%%%%%%%%%%%%%%%%%%%%%%

\begin{thebibliography}{10}
%%%%%%%%%%%%%%%%%%%%%%%%%%%%%%%%%%%%%%%%%%%%%%%%%%

\bibitem{Beringer:1900zz}
J.~Beringer et~al.
\newblock {Review of Particle Physics}.
\newblock {\em Phys. Rev. D}, 86:010001, 2012.

\bibitem{Mohler:2012nh}
Daniel Mohler.
\newblock {Review of lattice studies of resonances}.
\newblock {\em PoS}, LATTICE2012:003, 2012.

\bibitem{Luscher:1990ux}
M.~L{\"u}scher.
\newblock {T}wo-{P}article {S}tates on a {T}orus and {T}heir {R}elation to the
  {S}cattering{M}atrix.
\newblock {\em Nucl. Phys. B}, 354:531, 1991.

\bibitem{Luscher:1991cf}
M.~L{\"u}scher.
\newblock {S}ignatures of unstable particles in finite volume.
\newblock {\em Nucl. Phys. B}, 364:237, 1991.

\bibitem{Engel:2011aa}
Georg~P. Engel, C.~B. Lang, Markus Limmer, Daniel Mohler, and Andreas
  Sch{\"a}fer.
\newblock {QCD with two light dynamical chirally improved quarks: Mesons}.
\newblock {\em Phys. Rev. D}, 85:034508, 2012.

\bibitem{Engel:2011pp}
Georg~P. Engel, C.~B. Lang, Markus Limmer, Daniel Mohler, and Andreas
  Sch{\"a}fer.
\newblock {Excited meson spectroscopy with two chirally improved quarks}.
\newblock {\em PoS}, LATTICE2011:119, 2011.

\bibitem{Engel:2012qp}
Georg~P. Engel, C.~B. Lang, and Andreas Sch{\"a}fer.
\newblock {Low-lying {L}ambda {B}aryons {F}rom the {L}attice}.
\newblock {\em Phys. Rev. D}, 87:034502, 2013.

\bibitem{Engel:2013ig}
Georg~P. Engel, C.~B. Lang, Daniel Mohler, and Andreas Sch{{\"a}}fer.
\newblock {QCD with Two Light Dynamical Chirally Improved Quarks: Baryons}.
\newblock 2013.

\bibitem{Engel:2013ita}
Georg~P. Engel, C.~B. Lang, Daniel Mohler, and Andreas Sch{\"a}fer.
\newblock {QCD with Two Light Dynamical Chirally Improved Quarks}.
\newblock {\em Acta Phys.Polon.Supp.}, 6(3):873--878, 2013.

\bibitem{Bulava:2011np}
John Bulava.
\newblock {Progress on Excited Hadrons in Lattice QCD}.
\newblock {\em PoS}, LATTICE2011:021, 2011.

\bibitem{Lin:2011ti}
H.~W. Lin.
\newblock {Review of Baryon Spectroscopy in Lattice QCD}.
\newblock {\em Chin. J. Phys.}, 49:827, 2011.

\bibitem{Fodor:2012gf}
Zoltan Fodor and Christian Hoelbling.
\newblock {Light Hadron Masses from Lattice QCD}.
\newblock {\em Rev.Mod.Phys.}, 84:449, March 2012.

\bibitem{Ginsparg:1981bj}
P.~H. Ginsparg and K.~G. Wilson.
\newblock {A Remnant of Chiral Symmetry on the Lattice}.
\newblock {\em Phys. Rev. D}, 25:2649, 1982.

\bibitem{Luscher:1998pqa}
Martin L{\"u}scher.
\newblock {E}xact chiral symmetry on the lattice and the {G}insparg-{W}ilson
  relation.
\newblock {\em Phys. Lett. B}, 428:342, 1998.

\bibitem{Gattringer:2000js}
Christof Gattringer.
\newblock {A} new approach to {G}insparg-{W}ilson fermions.
\newblock {\em Phys. Rev. D}, 63:114501, 2001.

\bibitem{Gattringer:2000qu}
Christof Gattringer, Ivan Hip, and C.~B. Lang.
\newblock {A}pproximate {G}insparg-{W}ilson fermions: {A} first test.
\newblock {\em Nucl. Phys. B}, 597:451, 2001.

\bibitem{Luscher:1984xn}
M.~L{\"u}scher and P.~Weisz.
\newblock {O}n-shell improved lattice gauge theories.
\newblock {\em Commun. Math. Phys.}, 97:59, 1985.

\bibitem{Luscher:1990ck}
M.~L{\"u}scher and U.~Wolff.
\newblock {H}ow to calculate the {E}lastic {S}cattering {M}atrix in
  2-{D}imensional {Q}uantum{F}ield {T}heories by {N}umerical {S}imulation.
\newblock {\em Nucl. Phys. B}, 339:222, 1990.

\bibitem{Michael:1985ne}
C.~Michael.
\newblock {A}djoint {S}ources in {L}attice {G}auge {T}heory.
\newblock {\em Nucl. Phys. B}, 259:58, 1985.

\bibitem{Engel:2010my}
G.~P. Engel, C.~B. Lang, M.~Limmer, D.~Mohler, and A.~Sch{{\"a}}fer.
\newblock {Meson and baryon spectrum for QCD with two light dynamical quarks}.
\newblock {\em Phys. Rev. D}, 82:034505, 2010.

%%%%%%%%%%%%%%%%%%%%%%%%%%%%%%%%%%%%%%%%%%%%%%%%%%
\end{thebibliography}
\end{document}